\documentclass[sigconf,authorversion]{acmart}

\AtBeginDocument{%
  \providecommand\BibTeX{{%
    \normalfont B\kern-0.5em{\scshape i\kern-0.25em b}\kern-0.8em\TeX}}}

\copyrightyear{2021}
\acmYear{2021}
\setcopyright{acmcopyright}
\acmConference[AMSec '21]{Proceedings of the
2021 Workshop on Additive Manufacturing (3D Printing) Security}{November
19, 2021}{Virtual Event, Republic of Korea}
\acmBooktitle{Proceedings of the 2021 Workshop on Additive Manufacturing
(3D Printing) Security (AMSec '21), November 19, 2021, Virtual Event,
Republic of Korea}
\acmPrice{15.00}
\acmDOI{10.1145/3462223.3485619}
\acmISBN{978-1-4503-8480-3/21/11}

\usepackage{balance}

\usepackage{algorithm}
\usepackage{tikz}
\usepackage{fontawesome}
\usepackage[noend]{algpseudocode}
\newcommand{\FullReturn}{\State \textbf{return} }

\usepackage{todonotes} 

\newcommand{\drawrandom}{\xleftarrow\$}

\usepackage{pgfplots}

\newcommand{\id}{\mathsf{eID}}

\newcommand{\Hash}{\mathcal{H}}

\newcommand{\tx}{\mathsf{tx}}
\newcommand{\sk}{\mathsf{sk}}
\newcommand{\pk}{\mathsf{pk}}

\newcommand{\Setup}{\mathsf{Setup}}
\newcommand{\AccGen}{\mathsf{AccGen}}
\newcommand{\Spend}{\mathsf{Spend}}
\newcommand{\CoinGen}{\mathsf{CoinGen}}
\newcommand{\lts}{\mathsf{lts}}
\newcommand{\ltv}{\mathsf{ltv}}
\newcommand{\ltp}{\mathsf{ltp}}
\newcommand{\pp}{\mathsf{pp}}

\newcommand{\KDF}{\mathsf{KDF}}
\newcommand{\PK}{\mathsf{PK}}
\newcommand{\vsk}{\mathsf{vsk}}
\newcommand{\tsk}{\mathsf{tsk}}
\newcommand{\vpk}{\mathsf{vpk}}
\newcommand{\tpk}{\mathsf{tpk}}

\newcommand{\acct}{\mathsf{acc}}
\newcommand{\aux}{\mathsf{aux}}
\newcommand{\CS}{\mathsf{state}}
\newcommand{\ck}{\mathsf{ck}}

\newcommand{\seed}{\mathsf{seed}}
\newcommand{\bits}{{0,1}^*}

\newcommand{\ItemGen}{\mathsf{ItemGen}}
\newcommand{\OTGen}{\mathsf{OTGen}}
\newcommand{\Receive}{\mathsf{Receive}}
\newcommand{\CheckValue}{\mathsf{ChkVal}}
\newcommand{\CheckKey}{\mathsf{ChkKey}}

\newcommand{\View}{\mathsf{View}}
\newcommand{\Verify}{\mathsf{Verify}}
\newcommand{\typ}{\mathsf{ty}}

\newcommand{\sT}{\mathcal{T}}
\newcommand{\sS}{\mathcal{S}}

\newcommand{\cT}{{|\mathcal{T}|}}
\newcommand{\cS}{{|\mathcal{S}|}}

\newcommand*\circled[1]{\tikz[baseline=(char.base)]{
                \node[shape=circle,draw,inner sep=1pt] (char) {\small #1};}}

\begin{document}

\title{Confidential Token-Based License Management}

\author{Felix Engelmann}
\email{fe-research@nlogn.org}
\affiliation{%
\department{Department of Computer Science}
   \institution{Aarhus University}
   \city{Aarhus}
   \country{Denmark}
 }

 \author{Jan Philip Speichert}
\email{jan@idi-systems.com}
\author{Ralf God}
\email{ralf.god@tuhh.de}
\affiliation{
\department{Institute of Aircraft Cabin Systems}
   \institution{TUHH - Hamburg University of Technology}
   \city{Hamburg}
   \country{Germany}
   \postcode{21129}
 }
 
 \author{Frank Kargl}
\email{frank.kargl@uni-ulm.de}
\author{Christoph B{\"o}sch}
\email{christoph.boesch@uni-ulm.de}
\affiliation{
\department{Institute of Distributed Systems}
   \institution{Ulm University}
   \city{Ulm}
   \country{Germany}
 }

\begin{abstract}
  In a global economy with many competitive participants, licensing and tracking of 3D printed parts is desirable if not mandatory for many use-cases.
We investigate a blockchain-based approach, as blockchains provide many attractive features, like decentralized architecture and high security assurances.
An often neglected aspect of the product life-cycle management is the confidentiality of transactions to hide valuable business information from competitors.
To solve the combined problem of trust and confidentiality, we present a confidential licensing and tracking system which works on any publicly verifiable, token-based blockchain that supports tokens of different types representing licenses or attributes of parts.
Together with the secure integration of a unique $\id$ in each part, our system provides an efficient, immutable and authenticated transaction log scalable to thousands of transactions per second.
With our confidential Token-Based License Management system (cTLM), large industries such as automotive or aviation can license and trace all parts confidentially.
\end{abstract}



\begin{teaserfigure}
  \includegraphics[width=\textwidth]{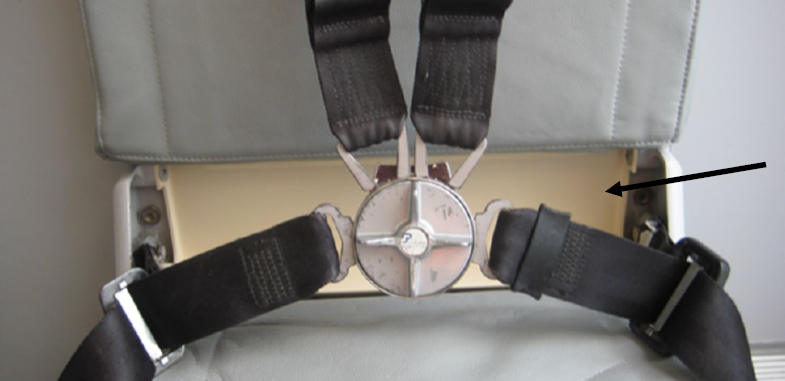}
  \caption{Installed belt mold spare part for the Airbus A300/A310 additively manufactured on demand.}
  \Description{Flight attendance seat with installed belt mold spare part.}
  \label{fig:belt_mold}
\end{teaserfigure}

\maketitle

\section{Introduction}
Driven by the current transformation in mechanical part design and manufacturing from product-centric processes to data-centric processes, both opportunities and challenges rise alike.
Technologies like additive layer manufacturing (ALM) enable an evolution from legacy part warehouses, logistic centers and just-in-time arrivals to a flexible manufacturing environment in which parts are produced on demand at or near the place of operation.
This decoupling of the physical location where a part is either designed or ordered from to the location where it is later manufactured has the potential to lower costs and speed up availability since, e.g., transportation costs and time are reduced to a minimum.
However, as it implies the transfer of highly sensitive design, development and production data into an unregulated environment, new challenges in terms of information security, intellectual property (IP) protection, and validation/verification of manufactured parts against the original design arise.
In order to establish an end-to-end chain of trust in such an ALM environment, every aspect from the design phase to the materialization of the actual part needs to be secured with an appropriate technology.

The identification of the part itself can be achieved easily by the integration of a tamper resistant electronic identification ($\id$) feature in the 3D-printed part as shown in~\cite{pagel18}, for example, by embedding an RFID chip.
Such an $\id$ feature is an enabler for identifying original parts, protecting intellectual property rights, granting licenses, and for tracking and tracing of parts in product life-cycle management systems (PLM-systems).
A monolithic integration of an $\id$ feature in e.g. a 3D-printed aircraft part provides a high level of physical security, effectively preventing a non-destructive removal of the $\id$ and therefore increasing the protection level against product piracy, fraud, and potential safety risks arising from non-authorized parts.

A real life use-case is an actual Airbus A300/A310 spare part which is only manufactured on demand due to unavailability of the original supplier and lack of legacy manufacturing tools.
The so-called belt mold (Figure~\ref{fig:belt_mold}) is part of the flight attendant seat and is currently manufactured with the aviation certified high performance thermoplastic ULTEM™.
Previous work~\cite{pagel18} already proved the practicability of integrating an $\id$ into the belt mold on an industrial grade 3D-Printer.

However, to realize a fully trusted supply chain it is necessary to establish data continuity from the first design of the 3D model to the actual print job at the print service provider, i.e. the company which operates the 3D printer and materializes the part.
Such a trusted supply chain enforces liability on each step and provides auditability.
As stated before, the described industrial environment is likely to be inhomogeneous as the print service provider may be able to produce parts for a variety of different industries.
While forwarding of actual printing data to printers can be established through traditional IT-systems, like webservers providing the data for authenticated download, the actual management and verification of licenses is more challenging.
The use of a blockchain-based approach is an attractive solution, as it omits the necessity of trust-centers and provides a high degree of data consistency and persistence for all participants.
The original design of blockchain technology is based on exhaustive transparency of all transactions stored in the blockchain in order to ensure their consistency and that data cannot be altered by anyone and unnoticed by the public.
In our use-case, this not only allows for a verification whether the printed part matches its design, but also for a licensing system where customer and contractor can handle their order management, adding a further level of security.
Together with the unique identification feature of the manufactured parts, our proposed system effectively reduces the risk of product piracy with a strict match between ordered parts and the $\id$s registered in the blockchain.

However, this transparency conflicts with the common approach how data is handled and controlled in a company, especially in industries where non-disclosure agreements (NDAs) are widely used in order to protect IP.
Storing the licensing data directly on a transparent blockchain violates such agreements and is seen as non-desirable or even prohibitive.
To many companies considering adopting blockchain-based systems, this problem is often not even clear during initial planing of projects.
To address this, we present our novel confidential Token-Based License Management (cTLM) which allows to satisfy both the requirement for a high level of data security in an open industrial environment as well as the requirement for a high level of privacy.
We achieve this licensing and tracking by representing licenses and attributes as tokens of different types.
In addition to persons and companies, parts are also identified by accounts in these systems.
The possession of a token of some type is then interpreted as having a specific property.
All tokens are transferred on a privacy-preserving Blockchain supporting multiple tokens, such as SwapCT~\cite{swapct}.

\section{Use Case Setting}

\begin{figure}
	\centering
	\begin{tikzpicture}[xscale=0.80,yscale=1]
		\node[align=center,text width=2cm] (design) at (0,0) {\color{red}\centering {\Huge\faUser}\\ Designer};
		\node[align=center,text width=2cm] (client) at (4,0) {\color{orange}\centering {\Huge\faUser}\\ Client};
		\node[align=center,text width=2cm] (printer) at (8,0) {\color{green!50!black}\centering {\Huge\faIndustry}\\ Printer};
		\node[align=center,text width=2cm] (object) at (8,-2) {\centering {\Huge\faCube}\\ Object};
		\draw[-latex, very thick] (design) .. controls (-0.5,-1.5) and (0.5,-1.5) .. node[below]{\circled{1}} (design);
		\draw[-latex, very thick] (design) -- node[below]{license} node[above]{\circled{2}} (client);
		\draw[-latex, very thick] (client) -- node[below]{order} node[above]{\circled{3}} (printer);
		\draw[-latex, very thick] (printer) -- node[right]{print} node[left]{\circled{4}} (object);
		\draw[-latex, very thick] (object) -- node[sloped,below]{deliver} node[above,sloped]{\circled{5}} (client);
		\draw[-latex,very thick, dashed] (client) -- node[right]{\faCube} node[left]{Life Cycle} +(0,-2.5);
	\end{tikzpicture}
	\caption{Setting of a license management system with entities and interactions}
	\label{fig:setting}
\end{figure}
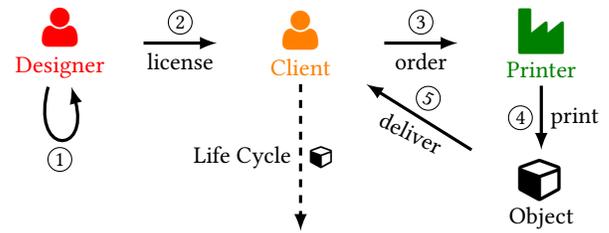

In this section, we describe the general setting in which we apply our license management system shown in Figure \ref{fig:setting}.
First, there is a designer creating a digital representation of a 3D object \circled{1}. This 3D file is then registered in our system by the designer, attributing this design to the designer. Once registered, the designer sells licenses either directly to manufacturers \circled{3} or to intermediaries \circled{2}. Anyone with a license can manufacture the part and register it by permanently binding the license to a unique identifier assigned to the part \circled{4}. Therefore parts are first-class citizens in our transaction system and have their own accounts to which license tokens are transferred. This allows someone handling the part to verify its legitimacy \circled{5}. Further attributes arising during the part's life cycle, e.g. assembly, disassembly, certification and recycling, are also permanently logged to allow for continuous tracking of the part.
This process flow is especially relevant in industries where parts need to be meticulously documented like aviation. As licensing and production data provides detailed insight and often leads to a competitive advantage, the actions must only be visible to entities involved in handling the parts.   

\section{Confidential Token Transaction System}
\label{sec:swapct}
Our system is built on top of an unspent transaction output (UTXO) protocol for type aware tokens. Generally, a transaction output is represented as a tuple $(a,\typ,P)$ of amount $a$ in type $\typ$ owned by party $P$. All three values are hidden inside a one-time account $\acct$.
A transaction in such a protocol take the form takes a list of previous transaction outputs which are used as inputs to the transaction. Their tokens are redistributed to a list of outputs.
The protocol has to assure that for each type, the sum of input amounts must be equal to the sum of output amounts and the input must belong to the transactor.
The protocol additionally has to provide a special transaction to create a new token type $\typ^*$ and assign its initial value $v^*$ to an account $P^*$.

Our System requires two types of long term accounts, one from which tokens are spendable and receive only accounts.
The regular long term accounts, e.g. $P^*$, have a secret key $\lts$ with $(\lts,\ltv,\ltp)\drawrandom\AccGen()$ and $\lts$ is used to authorize spending. Receive only accounts have no secret key and thereby are not be able to sign outgoing transactions $(\ltv,\ltp)\drawrandom \ItemGen(\id)$ where $\id$ is any identifier. A recipient retrieves all received tokens, from a read only account or a standard account with $\View(\ltv,\acct)$ where $\ltv$ is the long term view key available for regular account and receive only account.

\subsection{Concrete Scheme}

One possible privacy preserving transaction system with confidential types is SwapCT~\cite{swapct}.
It achieves sender and receiver anonymity by one-time addresses and stores amounts and types in commitments.

While SwapCT is focused on the conservation of tokens within one transaction, we formalize additional parts of the transaction system. I.e., $\Verify$ explicitly verifies a transaction in relation to a $\CS$ which captures all previous valid transactions. To support receive only accounts, we add an explicit $\ItemGen$ algorithm to generate an account without a spend key $\lts$. New token types are added to the system by $\CoinGen$ which generates a valid signature $\sigma$ which is accepted by $\Verify$. The SwapCT $\Receive$ algorithm is split into $\View$ and $\Receive$ to allow viewing of received tokens without requiring the spend key.
\begin{definition}
	A privacy-preserving multi-type system based on one-time accounts consists of the following algorithms which are a close adaptation of SwapCT:
	\begin{description}
		\item[$\pp\gets\Setup(1^\lambda)$:] takes the security parameter $\lambda$ and outputs public parameters $\pp$, implicitly given to the subsequent algorithms.
		\item[$(\lts,\ltv,\ltp)\gets\AccGen(\seed)$:] optionally takes a seed, if non provided uses a random seed and outputs a long term spend key $\lts$, view key $\ltv$ and the matching public key $\ltp$.
		\item[$(\ltv,\ltp)\gets\ItemGen(\id)$:] takes a part $\id\in\{0,1\}^\lambda$ and outputs a long term view key $\lts$ with matching public key $\ltp$.
		\item[$(\acct,\ck)\gets\OTGen(\ltp,\typ,a)$:] generates a one-time account from a long term public $\ltp$, a type $\typ\in\mathbb{T}$ and an amount $a$. It returns an account $\acct$ and a coin key $\ck$.
		\item[$\sigma\gets\Spend(\sS,\sT)$:] takes a set of inputs $\sS=\{a_i^\sS,\typ_i^\sS,\ck_i^\sS,\sk_i$, $\acct_i^\sS\}_{i=1}^\cS$ with amounts $a_i^\sS$ of typ $\typ_i^\sS$ with coin key $\ck_i^\sS$ and secret key $\sk_i$ for the account $\acct_i^\sS$. The outputs $\sT=(\ck_i^\sT,a_i^\sT,\typ_i^\sT,\acct_i^\sT)$ are defined by their amount $a_i^\sT$ of type $\typ_i^\sT$ hidden by coin key $\ck_i^\sT$ in the account $\acct_i^\sT$. The algorithm outputs a signature $\sigma$ for the transaction $\tx$ which is defined as $\tx(\sS,\sT):=(\{\aux_i^\sS\}_{i=1}^\sS,\{\acct_i^\sT\}_{i=1}^\cT)$ with some auxiliary information $\aux_i$ anonymously referencing the inputs.
		\item[$\sigma\gets\CoinGen(\sT)$:] takes a set of outputs defined as $\sT=(\ck_i^\sT$, $a_i^\sT$, $\typ_i^\sT,\acct_i^\sT)$ and outputs a signature $\sigma$ for the token initiation transaction defined as $\tx(\sT):=(\{\},\{\typ_i^\sT$, $\acct_i^\sT\}_{i=1}^\cT)$.
		\item[$\CS'/\bot\gets\Verify(\CS,\tx,\sigma)$:] takes the state $\CS$ of the system and a transaction $\tx$ with its signature $\sigma$. It either outputs a new state $\CS'$ with the transaction included or $\bot$.
		\item[$(a,\typ,\ck)\gets\View(\ltv,\acct)$:] takes a long term view key $\ltv$ for an account $\acct$ and outputs the containing amount $a$ of typ $\typ$ with coin key $\ck$.
		\item[$\sk\gets\Receive(\lts,\acct)$:] takes a long term secret key $\lts$ for an account $\acct$ and outputs the corresponding account secret key $\sk$.
		\item[$0/1\gets\CheckKey(\sk,\acct)$:]takes an account $\acct$ and a secret key $\sk$ and outputs a bit depending on the validity.
		\item[$0/1\gets\CheckValue(\ck,\typ,a,\acct)$:] takes an account $\acct$ a coin key $\ck$ and checks if the $\typ$ and amount $a$ match the account.
	\end{description}
\end{definition}

The system has to fulfill the following correctness criteria:
\begin{definition}[Correctness]
	A multi-type transaction system is correct if
	\begin{description}
		\item[correctly generated one-time accounts are viewable:] ~ \\ For all $\typ\in\mathbb{T}$ and all $a\in\{0,\dots,2^{64}-1\}$ and $(\lts,\ltv,\ltp)\gets\AccGen()$  or $(\ltv,\ltp)\gets\ItemGen(\id)$ the account \\$(\acct,\ck)\gets\OTGen(\ltp,\typ,a)$ is viewable with $(a',\typ',\ck')$ $\gets\View(\ltv,\acct)$
		such that $(a,\typ,\ck)=(a',\typ',\ck')$.
		\item[correctly generated one-time accounts from $\AccGen$ is] ~\\ \textbf{receivable:} For all $\typ\in\mathbb{T}$ and all $a\in\{0,\dots,2^{64}-1\}$ and $(\lts,\ltv,\ltp)\gets\AccGen()$ the account \\
		$(\acct,\ck)\gets\OTGen(\ltp,\typ,a)$ is receivable with \\ $\sk\gets\Receive(\lts,\acct)$ such that $\CheckKey(\sk',\acct)=1$.
		\item[honestly generated transactions should verify:] For any tuple $\sS,\sT$ with the structure from $\Spend$, where
		\begin{itemize}
			\item $\forall i\in[\cS]: \CheckKey(\sk_i,\acct_i^\sS)=1$
			\item $\forall i\in[\cS]: \CheckValue(\ck_i^\sS,\typ_i^\sS,a_i^\sS,\acct_i^\sS)=1$
			\item $\forall i\in[\cT]: \CheckValue(\ck_i^\sT,\typ_i^\sT,a_i^\sT,\acct_i^\sT)=1$
			\item $\forall \typ\in\{\typ_i^\sT\}_{i=1}^\cT: \sum_{\{i:\typ_i^\sS=\typ\}} a_i^\sS=\sum_{\{i:\typ_i^\sT=\typ\}} a_i^\sT$
		\end{itemize} 
		holds and all input accounts $\acct_i^\sS$ are spendable in $\CS$, it holds for any proof $\sigma\gets\Spend(\sS,\sT)$  that \\
		$\Verify(\CS,\tx,\sigma)\neq\bot$ with $\tx=\tx(\sS,\sT)$.
		\item[new generated types are valid:] For all tuples $\sT$ where all $\typ_i^\sT$ are not previously registered in $\CS$ and $\forall i\in[\cT]: \CheckValue(\ck_i^\sT,\typ_i^\sT,a_i^\sT,\acct_i^\sT)=1$, it holds for any \\ $\sigma\gets\CoinGen(\sT)$ that $\Verify(\CS,\tx,\sigma)\neq\bot$ with $\tx=\tx(\sT)$.
		
	\end{description}
	
\end{definition}

\subsection{Construction for missing algorithms}
For the algorithms which have no equivalent in SwapCT, we present the construction in this section. $\Spend$ is the immediate combination of $\mathsf{Seal}(\mathsf{Offer}(\sS,\sT))$. To generate a receive only account, we us a key derivation function $\KDF:\bits\times D\to\mathbb{K}$ where $D$ is the domain of labels for what the key is used and $\mathbb{K}$ is the appropriate key space. We present $\ItemGen$ in Algorithm \ref{alg:itemgen}. The tracking and view long term secret keys $\tsk,\vsk$ are derived from the $\id$ and their corresponding public keys are generated with the public key function $\PK$. Importantly, the spending long term public key $\pk$ is directly generated by the $\KDF$ such that the private key is infeasible to calculate.

\begin{algorithm}
	\caption{$\ItemGen$}
	\label{alg:itemgen}
	\begin{algorithmic}
		\Require $\id$
		\State $\tsk\gets\KDF(\id,\mathtt{track})$
		\State $\vsk\gets\KDF(\id,\mathtt{view})$
		\State $\vpk\gets\PK(\vsk)$
		\State $\tpk\gets\PK(\tsk)$
		\State $\pk\gets\KDF(\id,\mathtt{own})$
		\State $\ltp=(\vpk,\tpk,\pk)$
		\State $\ltv=(\vsk,\tsk)$
		\Return $\ltv,\ltp$
	\end{algorithmic}
\end{algorithm}

\subsection{Our requirements on a private transaction System}

If the above mentioned operations are supported in a privacy-preserving manner, the protocol is suitable for our application. We require the following privacy guarantees from the token protocol, for our system to be secure:
\begin{description}
 \item[Sender-Anonymity:] The owners of transaction inputs must only be known to the transaction creator.
 \item[Recipient-Anonymity:] The owners of outputs of a transaction must only be known to the sender, i.e. the creator of the transaction.
 \item[Confidentiality:] The amounts and types of tokens transferred in a transaction must only be known to the sender. Each recipient should only have access to the amount and type of their respective output.
\end{description}

\subsection{Generic Operations}

As most operations of our proposed system require a transfer of tokens or issuing a new token type, we present high level operations.

Given a state, an amount $a$ for a new type with domain $d\in\bits$ and pre-image $p\in\bits$ to the owner of the long term key $\ltp$ we issue a token type with Algorithm \ref{alg:issue}. We use a random oracle $\Hash:D\times\bits\to\mathbb{T}$ from the set of domains $D$ and arbitrary input to a type.

\begin{algorithm}
\caption{Issue Token}
\label{alg:issue}  
\begin{algorithmic}
\Require $\CS,a,d,p,\ltp$
\State $\typ \leftarrow \Hash(d\|p)$
\State $(\acct,\ck) \leftarrow \OTGen(\typ,a,\ltp)$
\State $\sigma \leftarrow \CoinGen(\sT=\{(\ck,a,\typ,\acct)\})$

\Return $\Verify(\CS,\tx(\sT),\sigma)$
\end{algorithmic}
\end{algorithm}

Given a state, an amount $a$ of tokens of typ $\typ$ from a one-time account $\acct$ with keys $(\lts,\ltv,\ltp)$, Algorithm \ref{alg:transfer} transfers the tokens to the owner of the long term key $\ltp^\sT$.

\begin{algorithm}
\caption{Transfer Token}
\label{alg:transfer}
\begin{algorithmic}
\Require $\CS,a,\typ,\acct^\sS,(\lts^\sS,\ltv^\sS,\ltp^\sS),\ltp^\sT$
\State $\sk\gets\Receive(\lts^\sS,\acct^\sS)$
\State $(\typ',s,\ck)\gets\View(\ltv^\sS,\acct^\sS)$
\If{$\typ'\neq\typ \lor s<a$} \Return $\bot$ 
\EndIf
\State $(\acct^\sT,\ck^\sT)\gets\OTGen(\typ,a,\ltp^\sT)$
\State $\sS=\{(s,\typ,\ck,\sk, \acct^\sS)\}$
\If{$s-a=0$}
\State $\sigma\gets\Spend(\sS,\sT=\{(\ck^\sT,a,\typ,\acct^\sT)\})$
\Else
\State $(\acct_R,\ck_R)\gets\OTGen(\typ,s-a,\ltp^\sS)$
\State $\sT=\{(\ck^\sT,a,\typ,\acct^\sT),(\ck_R,s-a,\typ,\acct_R)\}$
\State $\sigma\gets\Spend(\sS,\sT)$
\EndIf
\Return $\Verify(\CS,\tx(\sS,\sT),\sigma)$
\end{algorithmic}
\end{algorithm}

\section{Our Token-Based Licensing System}

Given the scenario above and a multi-type confidential transaction protocol as explained before, our system requires to track licenses and the state of objects in a distributed system. The persisted information must be immutable and should be visible only to those parties concerned. To achieve the persistence and immutability without a trusted third party, we use a distributed ledger.

An easy approach for a blockchain based license transaction system is to store the state of the licenses and tracked objects in a smart contract and programmatically update their state according to the smart contract rules. We show that a license management system does not require arbitrary state changes, as is possible with e.g. Ethereum, but can work on a much simpler transaction model. Our approach achieves this using a transaction logic for multiple token types. The benefit of this reduced requirement is an enhanced level of privacy. We propose an application on a blockchain based system to confidentially store the license and life-cycle information of parts which relies on distributed trust and zero knowledge proofs achieving a publicly verifiable history of operations. Therefore, we represent licenses and attributes of objects as tokens. Each 3D model is linked to its own type. Every token of this type is a license. Abstract attributes are types too and the tokens of this type certify that their owner holds the attribute. All tokens are then transferred confidentially. To assign a license or attribute to an object, the according tokens are transferred to the account of the object. Access to the $\id$ of a part allows reconstruction of the part's account for the verification of its attributes. In the remainder of the work, we use a multi-type system conforming to the requirements of Section~\ref{sec:swapct}.

The following sections describe the actions of the participants in more detail. First a new design is created and issued. Then license tokens of this design are traded. Once the part is manufactured, it gets a unique identifier and an account in the system. By transferring a license to the part it is permanently registered to the system. Any further actions in the life-cycle of the part, such as post processing or quality assurance, are then logged by attributes bound to the part's account. The actions are verifiable by anyone with access to the unique identifier. 

\subsection{Design Issuance}
The process starts with a designer creating a 3D object as a CAD file. The designers and engineers or their companies, respectively, own the intellectual property of the object's design. As our goal is to manage licenses of such designs, we have to identify the first and original owner and prevent others from claiming the property rights after the first registration. We identify a design by a cryptographic hash value of its CAD file. 

Small adaptations of the CAD file lead to different object identities, but the exact same file always results in the same identity.
Automatically assessing the similarity of two designs is difficult and out of scope of our work. Various authors \cite{hilaga2001topology,chen2017research,mun2011knowledge,lu2016selecting} proposed different approaches of similarity measures, all ineffective in a malicious setting due to heuristic approaches. In our setting the designers are malicious entities trying to claim ownership of an existing design of another designer.

\newcommand{\IssueToken}{\mathsf{IssueToken}}
\newcommand{\TransferToken}{\mathsf{TransferToken}}

\newcommand{\IssueDesign}{\mathsf{IssueDesign}}
\newcommand{\ApplyTransient}{\mathsf{ApplyTransient}}
\newcommand{\IssueCertificateToken}{\mathsf{IssueCertificateToken}}
\newcommand{\AttestPostProcessing}{\mathsf{AttestPostProcessing}}
\newcommand{\RegisterItem}{\mathsf{RegisterItem}}
\newcommand{\ItemVerification}{\mathsf{ItemVerification}}
\newcommand{\RecoverTransient}{\mathsf{RecoverTransient}}

The designer derives a new token type \[F_{\Box}=\Hash(\texttt{'design'}\|\mathtt{file.cad})\] and registers it in our system with an initial amount as a new token on the ledger. The designer now owns the initial supply of tokens of type $F_{\Box}$.
This initial supply should be chosen large enough because no new tokens of the same type can be generated again. For general parts, we suggest to have an initial supply of e.g. $2^{64}-1$ which is plentiful. On the other hand, artificial scarcity is achieved by, e.g., issuing only $100$ tokens. The designer is now in possession of the newly issued tokens. A transfer requires the matching private key of the designer. For all other participants, the registration process reveals only that someone registered a token type. Everyone can verify that this type was not registered before but gains no additional information.

Algorithm \ref{alg:register} shows the steps to register a design and requires an amount $a$, by default $a=2^{64}-1$, the file and the long term public key of the designer $\ltp_D$ generated by $\AccGen$. The domain \texttt{'design'} is used to separate tokens representing a CAD design from other token types in the system.

\begin{algorithm}
\caption{Issue Design}
\label{alg:register}
\begin{algorithmic}
\Require $\CS,a,\mathtt{file.cad},\ltp_D$

\Return $\IssueToken(\CS,a,\texttt{'design'},\mathtt{file.cad},\ltp_D)$
\end{algorithmic}
\end{algorithm}

\subsection{ Trading}
For any registered design, the designer initially owns all license tokens. Each token represents a license that allows the owner to manufacture one physical object. Most of the designers are not manufacturing the objects themselves but rather sell licenses to manufacturing companies, e.g. 3D print services. License tokens are therefore traded for other tokens in the system, most likely (but not limited to) representing fiat currencies. 

To exchange tokens of different types when using SwapCT, it provides a swap mechanism which ensures atomic execution. Either both parties receive each other’s tokens, or the swap is not performed at all. The exchanged tokens can either be used to manufacture the object or be traded again by performing another swap transaction. Each owner of a license token can use it to manufacture and register a legitimate object of this type.

The representation of a license as a token deviates slightly from a license in most legal systems. Tokens of a specific type are indifferentiable from each other and not batched together in purchases. Another difference is that there might be many intermediary owners of the license token, before it reaches the manufacturer which is uncommon in legal licensing contracts. Given such intermediary owners and due to the sender anonymity, the reconstruction of a transaction path back to the designer requires all previous owners to cooperate. Nevertheless, every participant of the system can verify that the total amount of tokens per type stays constant.

In a multi-type system which supports additional attributes, it is possible to implement an additional attribute which specifies the age of a token. The conservation non-interactive zero-knwoledge proof of each transaction then assures that all output tokens are one epoch older than the inputs. Thereby the token owner can verify the number of intermediaries.

\subsection{Manufacturing}
A manufacturer owning a license for an object can download the corresponding file from the designer or any shared storage. Our system implicitly protects the integrity of the file by linking it to the token type of the license. To verify the integrity, the file is hashed to a type $F'=\Hash(\texttt{'design'}\|\mathtt{file.cad})$ and then compared to the token type $F'\stackrel{?}{=}F_{\Box}$. If the two hash values match, the fetched file is correct and can be printed.

In most cases, the CAD file is fed into a CNC mill or a 3D printer to create the object.
After a successful object of the design exists, the manufacturer is trusted to, e.g., either measure (microscopic surface structure patterns, physically unclonable functions) or assign a unique identifier $\id$ to this part. Ideally, this $\id$ is integrated monolithically in the part in order to prevent an undesirable non-destructive extraction or alternation of the $\id$. A typical representation of an electronic identification feature is a radio-frequency identification (RFID) tag. RFID tags are widely used in a variety of branches like logistics, commerce, and identification documents. 

A minimal RFID system consists of a tag and a reader, where both tag and reader can either be a passive or an active component. The most typical setup for our setting is the Active Reader Passive Tag (ARPT) system \cite{manaffam2016rf}. The reader transmits an electromagnetic (EM) field which is used to send both signals and energy. The energy from the EM field induces a current in the antenna of the tag. This current is sufficient enough to power the tag's small integrated circuit (IC). Depending of the programming of the IC, the tag usually sends back a series of numbers, e.g. a simple $\id$ or a cryptographic signature. This allows for long-term, low-cost, and maintenance free identification system as the tag itself does not need any battery supply.
From a security perspective, the $\id$ must not be guessable for an object. According to the NIST SP800 \cite{dworkin2007recommendation} publication, this is achieved by a random bit-string of at least 96 bit. More importantly the $\id$ needs to be a unique feature of the part and difficult to copy to another part. For high stake parts, where identification is important, an integrated smart card might be used \cite{rankl2004smart}. The smart card has the benefit of generating a unique random private key and supports authentication challenges to prove the authenticity. This is much harder to copy onto another part as an RFID identifier, but is orders of magnitude more expensive.

From the $\id$, the manufacturer generates a new read only account $(\ltv,\ltp)=\ItemGen(\id)$.
This enables anyone knowing the $\id$ to verify which tokens were sent to the part by $\View(\ltv)$. To identify this part as legitimately manufactured, the license token has to be transferred to the part's account. The special feature of such an account, generating an immutable log of properties (received tokens), is the absence of a spend key.
Algorithm \ref{alg:registerpart} denotes the steps to register a valid part with an $\id$, the CAD file and the manufacturer's funding account $\acct_M$ along with the long term secret keys $(\lts_M,\ltv_M,\ltp_M)$.

\begin{algorithm}
\caption{Register Item}
\label{alg:registerpart} 
\begin{algorithmic}
\Require $\CS,\id,\mathtt{file.cad},\acct_M,(\lts_M,\ltv_M,\ltp_M)$
\State $\typ \leftarrow \Hash(\texttt{'design'}\|\mathtt{file.cad})$
\State $(\ltv_I,\ltp_I)=\ItemGen(\id)$

\FullReturn $\TransferToken(\CS,1,\typ,\acct_M,(\lts_M,\ltv_M,\ltp_M),\ltp_I)$
\end{algorithmic}
\end{algorithm}

\subsection{Post Processing \& Quality Assurance}
After manufacturing, parts regularly undergo post-processing and inspections. To attest such processes in the history of the part, the testing entity transfers a token to the part.

These tokens are not license tokens, but behave the same and can therefore use the same underlying confidential transactions. A property of parts is also represented in a new token type. Instead of the hash value of a CAD file, the new type is derived from hashing a unique string, e.g. \texttt{"Quality Assured by A. Inc."} in the domain \texttt{'property'}. The creator should publish this unique string to allow others to verify if a token really originated from them. The new type $F_{Q}$ is then registered on the ledger by Algorithm \ref{alg:cert} taking the unique string $Q$, the authorities $\ltp_Q$ and a large initial amount $a$. Every time the company decides to tag a part with this property, they transfer one token of the type to the part's account. 

\begin{algorithm}
\caption{Issue Certificate Tokens}
\label{alg:cert}
\begin{algorithmic}
\Require $\CS,a,Q,\ltp_Q$

\Return $\IssueToken(\CS,a,\texttt{'property'},Q,\ltp_Q)$
\end{algorithmic}
\end{algorithm}

The part then has the license token and the $F_Q$ token of the manufacturer. The absence of a private key for the part locks these tokens forever to the part. Thus, the tokens cannot be secretly transferred to a different part, given the unique identifier cannot be copied or cloned.

This process is repeated, if required, to log multiple properties to the part. On publication of a transfer, the transaction is timestamped persisting the date and time when the part received the property. The transaction to a part's account is not limited to simple binary properties. Small to medium scalar values can be stored by transferring multiple tokens of the same type to the object. Algorithm \ref{alg:post} requires the part $\id$, the type of certificate $Q$ and the spending key $\lts_Q$ of the owner of $\acct_Q$. The amount of tokens $a$ transferred represents the value. An example could be the quality of the part on a scale from one to five, depending on the number of $Q$ tokens received.
Instead of scalar values, the number of tokens can be interpreted as a map to strings.

\begin{algorithm}
\caption{Attest Post-Processing}
\label{alg:post}
\begin{algorithmic}
\Require $\CS,\id,Q,a=1,\acct_Q,(\lts_Q,\ltv_Q,\ltp_Q)$

\State $\typ \leftarrow \Hash(\texttt{'property'}\|Q)$
\State $(\ltv_I,\ltp_I)=\ItemGen(\id)$

\FullReturn $\TransferToken(\CS,a,\typ,\acct_Q,(\lts_Q,\ltv_Q,\ltp_Q),\ltp_I)$
\end{algorithmic}
\end{algorithm}

\subsection{Verification}
Anyone with physical access to the part can retrieve the $\id$ from the object and derive the account key with the same steps as the manufacturer previously did. With this key, all tokens transferred to the part can be recovered from the public ledger. The license token is checked to verify the original designer, $Q$ tokens attest a correct print and quality check by a known manufacturer and any further properties. This is shown in Algorithm \ref{alg:verify} which requires the parts $\id$ and recovers all received tokens. The transactions of the tokens are all timestamped, which allows the reconstruction of the part's history.

\begin{algorithm}
\caption{Item Verification}
\label{alg:verify}
\begin{algorithmic}
\Require $\CS,\id$
\State $t=\emptyset$
\State $(\ltv_I,\ltp_I)=\ItemGen(\id)$
\ForAll{$\acct\in\CS$}
\State  $d\gets\Receive(\ltv_I,\acct)$
\If{$d\neq\bot$}
        \State parse $d$ as $(\typ,a,\ck)$
    \State $t=t\cup(\typ,a,\ck)$
\EndIf
    \EndFor

\FullReturn $t$
\end{algorithmic}
\end{algorithm}

All digital systems are only as reliable as the information inserted into it. The interface between the physical world and the immutable ledger has to be trusted. At least our system incentivizes actors to act correctly, as an immutable documentation proves their operations in case of dispute.

For all other participants of the system without the part or the part number ($\id$), it is impossible to identify which tokens the part possesses. Thereby, the information is available to parties with legitimate interests only.

Our system can be a basis for a complete tracking and tracing system of parts which is often desired by industries, especially in the aviation industry, where a life-long tracking of parts is not only done in production but also during the full life-cycle of a part. This can easily be used to establish a real digital twin not only on system level, but on part level. While there are already several tracking and tracing systems on part level in aviation industry, none of them presents a complete end-to-end solution as our proposed system would.

\subsection{Proxy Accounts for Sensitive Parts}
\label{sec:proxy}
For sensitive parts, it might be interesting to limit the scope of which parties have insight into the tokens owned by the part.
While the ownership of tokens is directly derived from the $\id$ and every token received by the part is not spendable, we provide an extension where tracking and viewing of tokens is limited to a part owner. In this case, the view secret key $\ltv$ is not derived from the $\id$ directly but randomized by the owner by concatenating randomness $s$. For verification, the owner creates a designated verifier proof that the item received a specific amount of a given type without disclosing possible other tokens of the part. For a designated verifier party with key $\ltp_V$, the part owner creates a zero knowledge proof of knowledge (PoK):
\begin{align*}
\mathsf{P}&\mathsf{oK}\bigl(\mathsf{stmt}=(\acct,\ltp_V,\typ,a,\id) \exists \mathsf{wit}=(\lts_V,\ck,s)\\
&\text{ s.t. }\ItemGen(\id\|s)=(\ltv_I,\ltp_I) \land \View(\ltv_I,\acct) = (\ck,\typ,a)\\
&\phantom{ s.t. }\lor \ltp_V=\PK(\lts_V) \bigr) 
\end{align*}

Such a proof transcript still reveals the one-time account $\acct$ belonging to the part. To improve privacy at the cost of performance, the proxy party creates an anonymous reference with an anonymity set of $n$ accounts $\{\acct_i\}_{i=1}^n$:
\begin{align*}
\mathsf{P}&\mathsf{oK}\bigl(\mathsf{stmt}=(\{\acct_i\}_{i=1}^n,\ltp_V,\typ,a) \exists \mathsf{wit}=(\lts_V,j\in\{1,\dots,n\},\ck)\\
&\text{ s.t. }\ItemGen(\id\|s)=(\ltv_I,\ltp_I) \land \View(\ltv_I,\acct_j) = (\ck,\typ,a)\\
&\phantom{ s.t. }\lor \ltp_V=\PK(\lts_V) \bigr) 
\end{align*}
This does not reveal to the verifier which transaction output contains the license or property.

\subsection{Life cycle}
The digital representation of properties in an immutable ledger is especially relevant during the manufacturing process, but also provides multiple benefits during the life-cycle of the part. Parts being resold or refurbished are tracked by a token of a new type. As an example, B. Inc.'s business model is to buy old parts, refurbish them and then sell them again. This is realised with a token of type \texttt{"refurbished by B. Inc."}, which is transferred to the refurbished part. Everyone trusting that this token is really from B. Inc. can then check if a part is genuine and can view the history of production. This is again performed by Algorithm \ref{alg:cert} and \ref{alg:post}.

The identifier of the type belonging to known companies might be signed in a traditional PKI based system of which the root is controlled by a regulatory authority of the industry sector. E.g., aviation authorities such as the European Union Aviation Safety Agency (EASA) can sign token types from Original Equipment Manufacturer (OEM) or Tier-1 Suppliers in order to verify the originality of parts or services such as a refurbishment.

Once a property token is transferred to a part, it is impossible to take it away due to the lack or a private key. If removing tokens is desired, a simple convention can resolve this, by issuing a secondary token type, which negates the original one. Any part with only the positive token is valid, but if the part also owns the negated one, it is invalid. Once negated, no one can remove the negated token from the part. One example for such a negated token is the revocation of a part. Assume a company's inventory of parts get stolen. In this scenario, the company can issue a revocation token type \texttt{"revoked by A. Inc."} and transfer tokens to the stolen parts. Everyone trusting A. Inc. can check whether the part was revoked. Such revocation tokens have to be carefully verified if they are genuine or not, as they enable a denial of service attack, if such a token is transferred to a still working part.

\subsection{Transient Properties}
For many tokens owned by parts, it is crucial that they cannot be transferred away from the part to a different account belonging to a different part. This would facilitate the theft of license tokens and certification of subpar parts. However, some life-cycle properties are transient and being passed around between the part and people or machines interacting with it. This can e.g. be used for a per item payment system. Each part processed earns one token which is then used to claim a reward.

For this requirement, participants must be able to transfer tokens away from the object. This is achieved by a secondary account of each item with a spend key $P=\AccGen(\id)$. A set of $a$ tokens of type $T$ is applied to a part $\id$ from some operator with private key $\lts_O$ from account $\acct_O$ with Algorithm \ref{alg:applyt}. 
\begin{algorithm}
\caption{Apply Transient}
\label{alg:applyt}
\begin{algorithmic}
\Require $\CS,\id,T,a,\acct_O,(\lts_O,\ltv_O,\ltp_O)$

\State $\typ \leftarrow \Hash(\texttt{'attribute'}\|T)$
\State $(\ltv_I,\lts_I,\ltp_I)=\AccGen(\id)$

\FullReturn $\TransferToken(\CS,a,\typ,\acct_O,(\lts_O,\ltv_O,\ltp_O),\ltp_I)$
\end{algorithmic}
\end{algorithm}

All tokens held in this account can be claimed from the object by anyone having access to the $\id$, the type $T$, an amount $a$ (mostly $a=1$) and a recipient operator $\ltp_O$ with Algorithm \ref{alg:recovert}. Even after a token was transferred away from the part, it is still possible to verify that the object once owned the property token. The receiving and removal transactions even leave timestamps for both operations.
\begin{algorithm}
\caption{Recover Transient}
\label{alg:recovert}
\begin{algorithmic}
\Require $\CS,\id,a,\acct_I,\ltp_O$

\State $(\ltv_I,\lts_I,\ltp_I)=\AccGen(\id)$
\State $\sk\gets\Receive(\lts_I,\acct_I)$
\State $(\typ,s,\ck)\gets\View(\ltv_I,\acct_I)$

\FullReturn $\TransferToken(\CS,a,\typ,\acct_I,(\lts_I,\ltv_I,\ltp_I),\ltp_O)$

\end{algorithmic}
\end{algorithm}

\section{Security Analysis}

The confidentiality of the transaction data is important for the participants. Transfers of licenses quite always reflect some business interactions, which gives competitors valuable insight into metrics private to a company. The public nature of a blockchain requires all transactions to be public, so that a consensus can be reached. Our approach uses a blockchain which reaches a consensus and allows to verify the conservation rules publicly without revealing the data. This is achieved by using privacy-preserving transactions \cite{swapct}. Each transaction includes non-interactive zero knowledge (NIZK) proofs to convince other participants of the system that the transactor created the transaction according to the conservation rules. These NIZK proofs are publicly verifiable and are used to reach consensus about the validity of a transaction.

\subsection{Attacker Model}

We assume all actors of the system are malicious with the exception that they honestly provide inputs from the physical world. Meaning they try to attack the transaction system but do not insert bogus measurements.
The requirement for honest interfaces cannot be solved by any purely digital system. However the persistent logging incentivizes all participants to report correct data as they can be held accountable retrospectively.

The verification and persistence of transactions is performed by the entirety of actors following a consensus protocol. We assume that the majority of actors are honest and we therefore model the verification party as honest-but-curious. 

\subsection{Security Reduction}
All operations of our system either create transactions or receive transaction outputs in a way that the security model of the underlying transaction system is taking care of. So regarding anonymity and confidentiality, our system provides the same guarantees as the used privacy-preserving transaction system. An efficient attacker to the security of our system can be directly used to break the security properties of the underlying UTXO transaction system.

To securely deploy our token licensing system, several components require special attention. In the following sections, we elaborate on specific areas of confidentiality, which are important to license transactions and life-cycle tracking.
\subsection{Revealing Transactions}
In some circumstances, transactors have to reveal past transactions to third parties or authorities. If the underlying ledger uses anonymity sets, where each transactor is hidden in a set of unrelated transactors and too many transactions in a system are publicly known, it gets more and more difficult to stay anonymous within the network of transactions. So revealing some selected transactions to business partners or regulatory entities does not interfere with the anonymity in the system. It is generally in the participants’ interest to keep their transactions secret from the general public and especially from competitors.

However, in some industries access to information needs to be granted to certain institutions or authorities. Example use-cases are regular audits to ensure compliance or an investigation following an accident. To achieve this, an escrow key of the involved companies is stored at a safe place. The escrow key cannot be abused to gain insights into transactions of other participants.

If it is sufficient for the auditing entity, the part owners may present zero knowledge proofs of compliance instead of revealing the secret keys. This is achieved equivalently to the proxy accounts from Section \ref{sec:proxy}.
\subsection{Item $\id$ Enumeration}
Another important issue to care about, are $\id$s of parts. As they are the seed of the public account keys, they must not be guessable by someone not owning or having access to the part. Otherwise the part's account can be revealed. Also enumeration attacks where parts have consecutive numbers can lead to attacks on the confidentiality of tokens sent to these parts. A reliably non guessable source as of NIST-SP800 \cite{dworkin2007recommendation} is to use 96 bits of uniformly random data. From this, a part account is derived which is not guessable. This keeps the tokens transferred to the part secret to the owners of the part only. As previously mentioned, another option is to integrate a smart card with a randomly generated private key. 

It is still important to know that everyone who had access to the part in its history is able to store the $\id$ and track future transactions of the part.

\section{Performance}
To evaluate the performance of the system we take a closer look at the operations required and estimate a number of transactions per second. Operations are defined as the transfer of tokens from one account to another. Going back to the example of the aircraft industry an airplane is made up of around 350,000 \cite{nbc10} to 6 million \cite{altfeld2016commercial} parts from hundreds of different suppliers and subsuppliers. All those parts could be integrated into our license management system. Airbus and Boeing manufacture around 800 airplanes per year \cite{airbus19}. This equals to around 8,87 operations per second for a company like Airbus. Taking into account other industries like car manufacturing the numbers of operations are significantly higher. A car is made up of around 7,000 \cite{altfeld2016commercial} to 15,000 \cite{klier2008really} parts and 97 Million cars and commercial vehicles were produced in 2017 \cite{oica18}. If every part is produced based on a token transfer, this results in around 46,000 operations per second. Extending to the multitude of manufacturing sectors where a confidential license management system could be integrated the potential number of operations could be manifold.

\begin{figure}
 \begin{tikzpicture}
 \begin{axis}[xlabel={\# inputs and outputs}, ylabel={Generation time in s}, legend style={at={(0.03,0.97)},anchor=north west},width=\columnwidth ,
 height=5cm,
 y label style={at={(axis description cs:0,.5)}},
  x label style={at={(axis description cs:0.5,.05)}},
 ]
  
  \input{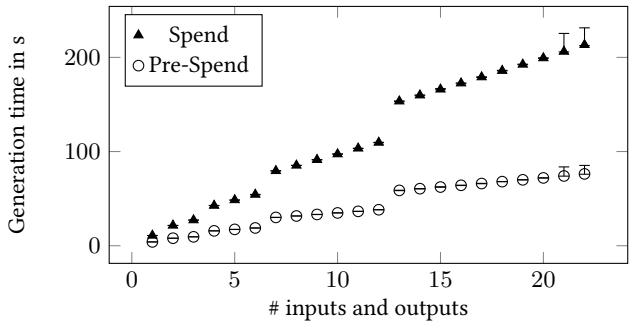}

 \end{axis}
 \end{tikzpicture}

 \caption{Transaction generation time of our prototype implementation on a Raspberry Pi first generation, representing an embedded system within a printer. As the secret keys are only required in part of the transaction, it is sufficient to execute the Pre-Spend on the embedded device and the remaining part can be offloaded to a powerful computer with minimal privacy loss. We use an anonymity set size of $27$. The error bars are the minimum and maximum runtime of 10 executions.}
 \label{fig:perf}
 \end{figure}

To show the applicability of our scheme, we evaluated the performance relevant for different scenarios. The transaction generation for e.g. registering items is most likely performed by the 3D printer. We use a Raspberry Pi first generation to represent the resource constraint device. Figure \ref{fig:perf} shows that usual transactions with few inputs and outputs require less than one minute, negligible in relation to the printing duration.

Transaction verification is most likely performed in an on-premise data center with powerful machines. As test payload we used transactions with 2 inputs and 2 outputs, as they will be most frequent in the system. On an dual socket AMD EPYC 7281 Server (64 threads) with Ubuntu 20.04, we achieved a median of 591 transaction verifications per second with a worst case performance of 582.

Another relevant time is scanning all transaction outputs for ones belonging to a part or an account. Only with the $\id$ it is possible to detect an output. We measured a median of 260,000 output verifications per second on the above mentioned server. 

From these measurements and an estimation of 50,000 operations per second and 5 kB storage per transaction, we derive the computational requirements of our system. Depending on the time by which a transaction output is reused in a subsequent transaction, the verification of transactions can be massively parallelized. For the verification to keep up with the transaction generation, each participant requires around 100 servers to verify all transactions in real time and one server to track new outputs. This is comparable to other systems with similar throughput. To store all transactions, around 20 TB per day are required. An efficient consensus mechanism easily achieves this throughput resulting in a global total order of transactions.

Within a trusted environment, e.g. in a single company, only one node is required to verify and store all transactions. Smaller nodes trusting the central entity can validate part properties based on the central, verified storage. For a system tracking each part of every car manufactured, these requirements are possible with current commodity hardware and allow an immutable, publicly verified log of all operations

\subsection{Concurrency}
Every state change of a part is reflected with a transaction. However, the order of the persisted transactions cannot be determined in advance. Transacting two tokens to the account of a part at approximately the same instant will result in an arbitrary order. There are no guarantees like first in first out, even for a single part, as each transaction is hiding the recipient and there might be a separate actor transmitting a token to said part. To implement strict conditional dependencies, it is necessary to query the persisted log, if the condition is persisted.

The certification accounts have to split their initial output, after issuing, into multiple smaller outputs. Using the change output from the previous part is only possible after the transaction is persisted, so repetitively using the change to attest many parts in short succession is not possible. This issue is mitigated by using smaller outputs in a round robin fashion.

\section{Related Work}
As our system combines privacy-preserving blockchain protocols with blockchain based license management, we present recent research results from both areas.

\subsection{Privacy Perserving Blockchains}

\paragraph{Cloaked Assets (ZkVM)}
The team of Stellar\footnote{https://www.stellar.org/} introduces a zero knowledge virtual machine \cite{zkvm}, which executes smart contracts in a privacy-preserving manner. This enables management of typed tokens and provides the necessary functionality to be used as a basis for our application.
We chose SwapCT as Cloaked Assets do not provide sender and receiver anonymity. 
\paragraph{Conficential Assets on MimbleWimble}
Yi Zheng et al. \cite{mimblewimbleca} extended the MimbleWimble protocol with confidential assets to support multiple tokens in a confidential way. This chain may be used as a basis for our system with the limitation, that transactions are created interactively with the recipients. The aggregation characteristics of MimbleWimble are highly favourable for high throughput applications at the cost of reduced privacy.

\subsection{Blockchain PLM/DRM} 
Multiple authors have described the use of blockchain technology for product life-cycle management \cite{abeyratne2016blockchain,engelmann2018intellectual,heber2017towards}. Unfortunately, all previously presented solutions neglect the crucial confidentiality of operations. With all transactions in public, valuable business metrics are analysed by everyone and may be used for harm. The alternative to deploy a private ledger in a consortium does not solve the confidentiality requirement, as the very competitors, to who insight is most valuable, still have access as they most likely belong to the same consortium. Our solution maintains the confidentiality and is therefore also suitable for private consortia as well as public ledgers, depending on the participants and the public interest.

Herbert and Litchfield \cite{bclicense} propose two blockchain based software license managements. The first uses additional semantics of standard bitcoin transactions, providing no confidentiality. In their second scheme, the vendor is trusted to manage encrypted licenses persisted within transactions. While improving data protection, they require an honest entity to achieve the same confidentiality as our protocol.

\section{Conclusion}
The tracking and tracing of parts in complex industries such as automotive or aviation is a fundamental task, especially having intricate business relations. In addition, reliably differentiating legitimately licensed parts from reputable sources is a difficult feat. On top, the information about parts must remain secret from competitors and alike. 

Our proposed system is able to provide all required functionality while maintaining confidentiality of operations in an untrusted environment. We defined protocols to represent the complex life-cycle actions in the form of basic transfers of tokens. By building the system on top of a privacy-preserving multi token blockchain, our system inherits the confidentiality guarantees. Given an efficient consensus mechanism for transaction ordering, the verification is massively parallelizable and can handle up to 10 transactions per core per second. These features combined with a decent performance makes our system directly applicable to many real world life-cycle management scenarios.

\section*{Acknowledgements}

This work was supported by the Federal Ministry of Economic Affairs and Energy on the basis of a decision by the German Bundestag.

\bibliographystyle{ACM-Reference-Format}
\balance
\bibliography{bib}


\begin{thebibliography}{20}


\ifx \showCODEN    \undefined \def \showCODEN     #1{\unskip}     \fi
\ifx \showDOI      \undefined \def \showDOI       #1{#1}\fi
\ifx \showISBNx    \undefined \def \showISBNx     #1{\unskip}     \fi
\ifx \showISBNxiii \undefined \def \showISBNxiii  #1{\unskip}     \fi
\ifx \showISSN     \undefined \def \showISSN      #1{\unskip}     \fi
\ifx \showLCCN     \undefined \def \showLCCN      #1{\unskip}     \fi
\ifx \shownote     \undefined \def \shownote      #1{#1}          \fi
\ifx \showarticletitle \undefined \def \showarticletitle #1{#1}   \fi
\ifx \showURL      \undefined \def \showURL       {\relax}        \fi
\providecommand\bibfield[2]{#2}
\providecommand\bibinfo[2]{#2}
\providecommand\natexlab[1]{#1}
\providecommand\showeprint[2][]{arXiv:#2}

\bibitem[\protect\citeauthoryear{Abeyratne and Monfared}{Abeyratne and
  Monfared}{2016}]%
        {abeyratne2016blockchain}
\bibfield{author}{\bibinfo{person}{Saveen~A Abeyratne} {and}
  \bibinfo{person}{Radmehr~P Monfared}.} \bibinfo{year}{2016}\natexlab{}.
\newblock \showarticletitle{Blockchain ready manufacturing supply chain using
  distributed ledger}.
\newblock  (\bibinfo{year}{2016}).
\newblock


\bibitem[\protect\citeauthoryear{Airbus}{Airbus}{2019}]%
        {airbus19}
\bibfield{author}{\bibinfo{person}{Airbus}.} \bibinfo{year}{2019}\natexlab{}.
\newblock \bibinfo{booktitle}{\emph{Airbus achieves new commercial aircraft
  delivery record in 2018}}.
\newblock
\urldef\tempurl%
\url{https://www.airbus.com/newsroom/press-releases/en/2019/01/airbus-achieves-new-commercial-aircraft-delivery-record-in-2018.html}
\showURL{%
\tempurl}


\bibitem[\protect\citeauthoryear{Altfeld}{Altfeld}{2016}]%
        {altfeld2016commercial}
\bibfield{author}{\bibinfo{person}{Hans-Henrich Altfeld}.}
  \bibinfo{year}{2016}\natexlab{}.
\newblock \bibinfo{booktitle}{\emph{Commercial aircraft projects: Managing the
  development of highly complex products}}.
\newblock \bibinfo{publisher}{Routledge}.
\newblock


\bibitem[\protect\citeauthoryear{Andreev, Glickstein, Niu, Rinearson, Sur, and
  Yun}{Andreev et~al\mbox{.}}{2019}]%
        {zkvm}
\bibfield{author}{\bibinfo{person}{Oleg Andreev}, \bibinfo{person}{Bob
  Glickstein}, \bibinfo{person}{Vicki Niu}, \bibinfo{person}{Tess Rinearson},
  \bibinfo{person}{Debnil Sur}, {and} \bibinfo{person}{Cathie Yun}.}
  \bibinfo{year}{2019}\natexlab{}.
\newblock \bibinfo{booktitle}{\emph{{ZkVM: fast, private, flexible blockchain
  contracts}}}.
\newblock \bibinfo{type}{{T}echnical {R}eport}.
\newblock


\bibitem[\protect\citeauthoryear{Chen, Hao, Liu, Han, and Ye}{Chen
  et~al\mbox{.}}{2017}]%
        {chen2017research}
\bibfield{author}{\bibinfo{person}{Xin Chen}, \bibinfo{person}{Jingbin Hao},
  \bibinfo{person}{Hao Liu}, \bibinfo{person}{Zhengtong Han}, {and}
  \bibinfo{person}{Shengping Ye}.} \bibinfo{year}{2017}\natexlab{}.
\newblock \showarticletitle{Research on Similarity Measurements of 3D Models
  Based on Skeleton Trees}.
\newblock \bibinfo{journal}{\emph{Computers}} \bibinfo{volume}{6},
  \bibinfo{number}{2} (\bibinfo{year}{2017}), \bibinfo{pages}{17}.
\newblock


\bibitem[\protect\citeauthoryear{Dworkin}{Dworkin}{2007}]%
        {dworkin2007recommendation}
\bibfield{author}{\bibinfo{person}{Morris~J Dworkin}.}
  \bibinfo{year}{2007}\natexlab{}.
\newblock \bibinfo{booktitle}{\emph{Recommendation for block cipher modes of
  operation: Galois/Counter Mode (GCM) and GMAC}}.
\newblock \bibinfo{type}{{T}echnical {R}eport}.
\newblock


\bibitem[\protect\citeauthoryear{Engelmann, Holland, Nigischer, and
  Stjepandi{\'c}}{Engelmann et~al\mbox{.}}{2018}]%
        {engelmann2018intellectual}
\bibfield{author}{\bibinfo{person}{Felix Engelmann}, \bibinfo{person}{Martin
  Holland}, \bibinfo{person}{Christopher Nigischer}, {and}
  \bibinfo{person}{Josip Stjepandi{\'c}}.} \bibinfo{year}{2018}\natexlab{}.
\newblock \showarticletitle{Intellectual property protection and licensing of
  3D print with blockchain technology}. In
  \bibinfo{booktitle}{\emph{Transdisciplinary Engineering Methods for Social
  Innovation of Industry 4.0: Proceedings of the 25th ISPE Inc. International
  Conference on Transdisciplinary Engineering, July 3--6, 2018}},
  Vol.~\bibinfo{volume}{7}. IOS Press, \bibinfo{pages}{103}.
\newblock


\bibitem[\protect\citeauthoryear{Engelmann, Müller, Peter, Kargl, and
  Bösch}{Engelmann et~al\mbox{.}}{2021}]%
        {swapct}
\bibfield{author}{\bibinfo{person}{Felix Engelmann}, \bibinfo{person}{Lukas
  Müller}, \bibinfo{person}{Andreas Peter}, \bibinfo{person}{Frank Kargl},
  {and} \bibinfo{person}{Christoph Bösch}.} \bibinfo{year}{2021}\natexlab{}.
\newblock \showarticletitle{SwapCT: Swap Confidential Transactions for
  Privacy-Preserving Multi-Token Exchanges}.
\newblock \bibinfo{howpublished}{Cryptology ePrint Archive, Report 2021/631}.
\newblock  (\bibinfo{year}{2021}).
\newblock
\newblock
\shownote{\url{https://eprint.iacr.org/2021/631}}.


\bibitem[\protect\citeauthoryear{Heber, Groll, et~al\mbox{.}}{Heber
  et~al\mbox{.}}{2017}]%
        {heber2017towards}
\bibfield{author}{\bibinfo{person}{Dominik Heber}, \bibinfo{person}{Marco
  Groll}, {et~al\mbox{.}}} \bibinfo{year}{2017}\natexlab{}.
\newblock \showarticletitle{Towards a digital twin: How the blockchain can
  foster E/E-traceability in consideration of model-based systems engineering}.
  In \bibinfo{booktitle}{\emph{DS 87-3 Proceedings of the 21st International
  Conference on Engineering Design (ICED 17) Vol 3: Product, Services and
  Systems Design, Vancouver, Canada, 21-25.08. 2017}}.
  \bibinfo{pages}{321--330}.
\newblock


\bibitem[\protect\citeauthoryear{Herbert and Litchfield}{Herbert and
  Litchfield}{2015}]%
        {bclicense}
\bibfield{author}{\bibinfo{person}{Jeff Herbert} {and} \bibinfo{person}{Alan
  Litchfield}.} \bibinfo{year}{2015}\natexlab{}.
\newblock \showarticletitle{A novel method for decentralised peer-to-peer
  software license validation using cryptocurrency blockchain technology}. In
  \bibinfo{booktitle}{\emph{Proceedings of the 38th Australasian computer
  science conference (ACSC 2015)}}, Vol.~\bibinfo{volume}{27}.
  \bibinfo{pages}{30}.
\newblock


\bibitem[\protect\citeauthoryear{Hilaga, Shinagawa, Kohmura, and Kunii}{Hilaga
  et~al\mbox{.}}{2001}]%
        {hilaga2001topology}
\bibfield{author}{\bibinfo{person}{Masaki Hilaga}, \bibinfo{person}{Yoshihisa
  Shinagawa}, \bibinfo{person}{Taku Kohmura}, {and} \bibinfo{person}{Tosiyasu~L
  Kunii}.} \bibinfo{year}{2001}\natexlab{}.
\newblock \showarticletitle{Topology matching for fully automatic similarity
  estimation of 3D shapes}. In \bibinfo{booktitle}{\emph{Proceedings of the
  28th annual conference on Computer graphics and interactive techniques}}.
  ACM, \bibinfo{pages}{203--212}.
\newblock


\bibitem[\protect\citeauthoryear{Klier and Rubenstein}{Klier and
  Rubenstein}{2008}]%
        {klier2008really}
\bibfield{author}{\bibinfo{person}{Thomas~H Klier} {and}
  \bibinfo{person}{James~M Rubenstein}.} \bibinfo{year}{2008}\natexlab{}.
\newblock \showarticletitle{Who really made your car?}
\newblock \bibinfo{journal}{\emph{Employment Research Newsletter}}
  \bibinfo{volume}{15}, \bibinfo{number}{2} (\bibinfo{year}{2008}),
  \bibinfo{pages}{1}.
\newblock


\bibitem[\protect\citeauthoryear{Lu, Qin, Qi, Zeng, Zhong, Liu, and Jiang}{Lu
  et~al\mbox{.}}{2016}]%
        {lu2016selecting}
\bibfield{author}{\bibinfo{person}{Wenlong Lu}, \bibinfo{person}{Yuchu Qin},
  \bibinfo{person}{Qunfen Qi}, \bibinfo{person}{Wenhan Zeng},
  \bibinfo{person}{Yanru Zhong}, \bibinfo{person}{Xiaojun Liu}, {and}
  \bibinfo{person}{Xiangqian Jiang}.} \bibinfo{year}{2016}\natexlab{}.
\newblock \showarticletitle{Selecting a semantic similarity measure for
  concepts in two different CAD model data ontologies}.
\newblock \bibinfo{journal}{\emph{Advanced Engineering Informatics}}
  \bibinfo{volume}{30}, \bibinfo{number}{3} (\bibinfo{year}{2016}),
  \bibinfo{pages}{449--466}.
\newblock


\bibitem[\protect\citeauthoryear{Manaffam and Jabalameli}{Manaffam and
  Jabalameli}{2016}]%
        {manaffam2016rf}
\bibfield{author}{\bibinfo{person}{Saeed Manaffam} {and}
  \bibinfo{person}{Amirhossein Jabalameli}.} \bibinfo{year}{2016}\natexlab{}.
\newblock \showarticletitle{RF-localize: An RFID-based localization algorithm
  for Internet-of-Things}. In \bibinfo{booktitle}{\emph{2016 Annual IEEE
  Systems Conference (SysCon)}}. IEEE, \bibinfo{pages}{1--5}.
\newblock


\bibitem[\protect\citeauthoryear{Mun and Ramani}{Mun and Ramani}{2011}]%
        {mun2011knowledge}
\bibfield{author}{\bibinfo{person}{Duhwan Mun} {and} \bibinfo{person}{Karthik
  Ramani}.} \bibinfo{year}{2011}\natexlab{}.
\newblock \showarticletitle{Knowledge-based part similarity measurement
  utilizing ontology and multi-criteria decision making technique}.
\newblock \bibinfo{journal}{\emph{Advanced Engineering Informatics}}
  \bibinfo{volume}{25}, \bibinfo{number}{2} (\bibinfo{year}{2011}),
  \bibinfo{pages}{119--130}.
\newblock


\bibitem[\protect\citeauthoryear{{NBC News}}{{NBC News}}{2010}]%
        {nbc10}
\bibfield{author}{\bibinfo{person}{Allison~Linn {NBC News}}.}
  \bibinfo{year}{2010}\natexlab{}.
\newblock \bibinfo{booktitle}{\emph{Hundreds of suppliers, one Boing 737
  airplane}}.
\newblock
\urldef\tempurl%
\url{http://www.nbcnews.com/id/36507420/ns/business-us\_business/t/hundreds-suppliers-one-boeing-airplane/\#.XTmO4ZMzZaQ}
\showURL{%
\tempurl}


\bibitem[\protect\citeauthoryear{of~Motor Vehicle~Manufacturers}{of~Motor
  Vehicle~Manufacturers}{2018}]%
        {oica18}
\bibfield{author}{\bibinfo{person}{International~Organization of Motor
  Vehicle~Manufacturers}.} \bibinfo{year}{2018}\natexlab{}.
\newblock \bibinfo{booktitle}{\emph{2017 Production statistics}}.
\newblock
\urldef\tempurl%
\url{http://www.oica.net/category/production-statistics/2017-statistics/}
\showURL{%
\tempurl}


\bibitem[\protect\citeauthoryear{Pagel, Speichert, and God}{Pagel
  et~al\mbox{.}}{2018}]%
        {pagel18}
\bibfield{author}{\bibinfo{person}{A Pagel}, \bibinfo{person}{Jan~Philip
  Speichert}, {and} \bibinfo{person}{Ralf God}.}
  \bibinfo{year}{2018}\natexlab{}.
\newblock \showarticletitle{{Wie sicher ist der 3D-Druck?}}
\newblock \bibinfo{journal}{\emph{Konstruktionspraxis}} \bibinfo{volume}{2018},
  \bibinfo{number}{10} (\bibinfo{year}{2018}), \bibinfo{pages}{34--46}.
\newblock


\bibitem[\protect\citeauthoryear{Rankl and Effing}{Rankl and Effing}{2004}]%
        {rankl2004smart}
\bibfield{author}{\bibinfo{person}{Wolfgang Rankl} {and}
  \bibinfo{person}{Wolfgang Effing}.} \bibinfo{year}{2004}\natexlab{}.
\newblock \bibinfo{booktitle}{\emph{Smart card handbook}}.
\newblock \bibinfo{publisher}{John Wiley \& Sons}.
\newblock


\bibitem[\protect\citeauthoryear{Zheng, Ye, Dai, Sun, and Gelfer}{Zheng
  et~al\mbox{.}}{2019}]%
        {mimblewimbleca}
\bibfield{author}{\bibinfo{person}{Yi Zheng}, \bibinfo{person}{Howard Ye},
  \bibinfo{person}{Patrick Dai}, \bibinfo{person}{Tongcheng Sun}, {and}
  \bibinfo{person}{Vladislav Gelfer}.} \bibinfo{year}{2019}\natexlab{}.
\newblock \bibinfo{title}{Confidential Assets on MimbleWimble}.
\newblock \bibinfo{howpublished}{Cryptology ePrint Archive, Report 2019/1435}.
\newblock
\newblock
\shownote{\url{https://eprint.iacr.org/2019/1435}}.


\end{thebibliography}

\end{document}